\documentclass[12pt]{article}
\usepackage{helvet,times,mathptm}
\usepackage{graphicx}
\begin{document}

\begin{center}
{\noindent {\bf Exact Entanglement Studies of Strongly Correlated Systems: 
Role of Long-Range Interactions and Symmetries of the System}}
\end{center}

\begin{center}
Shaon Sahoo,$^{a,}$\footnote[1]{shaon@physics.iisc.ernet.in} 
V. M. L. Durga Prasad Goli,$^{b,}$\footnote[2]{gvmldurgaprasad@sscu.iisc.ernet.in}
S. Ramasesha,$^{b,}$\footnote[3]{ramasesh@sscu.iisc.ernet.in} 
and  Diptiman Sen$^{c,}$\footnote[4]{diptiman@cts.iisc.ernet.in}
\end{center}

\begin{center}
{\small \it 
$^a$Department of Physics, Indian Institute of Science, Bangalore 560012, 
India\\
$^b$Solid State $\&$ Structural Chemistry Unit, Indian Institute of Science, 
Bangalore 560012, India\\
$^c$Centre for High Energy Physics, Indian Institute of Science, 
Bangalore 560012, India\\}
\end{center}

\begin{abstract}
{\noindent
We study the bipartite entanglement of strongly correlated systems using 
exact diagonalization techniques. In particular, we examine how the 
entanglement changes in the presence of long-range interactions by studying 
the Pariser-Parr-Pople model with long-range interactions. We compare the 
results for this model with those obtained for the Hubbard and Heisenberg
models with short-range interactions. This study helps us to understand why 
the density matrix renormalization group (DMRG) technique is so successful 
even in the presence of long-range interactions. To better understand the 
behavior of long-range interactions and why the DMRG works well with it, 
we study the entanglement spectrum of the ground state and a few excited 
states of finite chains. We also investigate if the symmetry properties of 
a state vector have any significance in relation to its entanglement. 
Finally, we make an interesting observation on the entanglement profiles 
of different states (across the energy spectrum) in comparison with the 
the corresponding profile of the density of states. We use isotropic 
chains and a molecule with non-Abelian symmetry for these numerical 
investigations.}
\end{abstract}

\section{Introduction}

Quantum entanglement, which began with the famous EPR controversy 
\cite{mermin}, has become one of the most exciting and important 
fields of research in recent times. Stated simply, entanglement is the 
quantum correlation among different parts of a many-body system. It is a
property of a quantum state and does not have any classical counterpart. 
It signifies that when a quantum system consists of two parts, one part 
is correlated with the other part even when the two parts
are far apart and not `physically' connected. Besides being fundamentally 
important in understanding or interpreting quantum mechanics, there are 
three main motivations for considering entanglement seriously 
\cite{vedral}; it can be used to study the quantum-to-classical transition, 
it can be used for studying quantum phase transitions, and it has the
potential to play a significant role in modern technology. Entanglement is at 
the heart of quantum computation and quantum communication which may be 
viewed as the future of computation and communication science.

In this paper, we will study the entanglement of a number of many-body 
systems in order to characterize and better understand the eigenstates of a 
many-body Hamiltonian. A wide range of entanglement studies have already been 
carried out on many-body systems \cite{amico}, but to the best of our 
knowledge, the role of long-range interactions has not been investigated. 
Unlike many of the other studies of many-body
systems, we will not try to find the scaling properties of entanglement with
the system size; instead, we will make a
comparative study of finite systems to understand the relative behaviors of 
different interacting models. The Pariser-Parr-Pople (PPP) Hamiltonian with 
standard parameters, the Hubbard Hamiltonian with different values of the 
on-site interaction and the antiferromagnetic Heisenberg Hamiltonian with
both spin 1/2 and 1 sites will be used for our comparative studies. The 
investigations will be done for the ground states of chains with either 
varying lengths (with equal block sizes) or varying block sizes (with fixed
chain length). In the case of the interacting electron systems, we will also
study a few important excited states such as $1^1A_g^+$, $2^1A_g^+$, 
$1^1B_u^-$ and $1^3B_u^+$. These correspond to the important states of 
a conjugated polymer, namely, the ground state
($1^1A_g^+$), the lowest energy two-photon state ($2^1A_g^+$), the lowest 
energy one-photon state ($1^1B_u^-$), and the lowest energy triplet state 
($1^3B_u^+$). The labels `$A$' and `$B$' correspond to even and odd parity 
states under reflection of the chain. For linear chains these
are also even or odd under inversion, and hence the subscript `$g$' goes 
with `$A$' state and `$u$' with `$B$' states. The superscript `+' or `-' 
refers to even or odd states under electron-hole symmetry. The `+' space 
includes `covalent' many-body basis states with one electron per site, 
while `ionic' space excludes these states. To better understand the effect 
of long-range interactions, we will also study the entanglement spectrum of 
the ground state of a chain. A comparison will be done with the results for 
the short-range interacting Hubbard model (with $U/t = 4$).

Our study serves the important purpose of understanding the high accuracy of
density matrix renormalization group (DMRG) technique \cite{schollwock} when 
applied to the PPP model, even though it incorporates long-range 
electron-electron interactions. 
The success of the DMRG method in low-dimensional systems is attributed to 
the connection between the area law of the entanglement entropy, $S$ 
\cite{srednicki,vidal}, and the cut-off, $m$, in the number of density matrix 
eigenvectors (DMEVs) that should be retained in a DMRG calculation for a 
given accuracy. The entanglement entropy of a system surrounded by an 
environment scales as the area between the system and the environment. The 
cut-off in the number of DMEVs for a fixed accuracy scales exponentially with 
the entanglement entropy ($m\sim 2^S$) \cite{schollwock}. 
However, one-dimensional systems away from criticality have a constant 
entanglement entropy independent of the size of the system. This explains why 
the DMRG method is so successful in studying correlated
one-dimensional systems. It also underlines the difficulties with the DMRG 
technique in studying higher dimensional systems as well as one-dimensional 
systems close to criticality; the latter systems have a logarithmic correction 
to the entropy which leads to an increase in the entanglement entropy
as one approaches the critical point. The area law arises due to the 
dangling bonds which are created by cutting the system from the environment. 
This would seem to imply that when we have systems with long-range
interactions, even in one dimension the DMRG method should not retain the 
same accuracy for a given cut-off as the system size is increased because 
the number of interactions that are cut resulting in
dangling bonds increases with increasing system size. However, DMRG 
calculations on quasi-one-dimensional systems with diagonal long-range 
interactions have proved to be very accurate. Our entanglement study of
one-dimensional systems is aimed at understanding this feature.

We will also investigate the entanglement properties of a 12-site icosahedral 
system using the PPP Hamiltonian. This system has the largest non-Abelian 
point group symmetry; this enables us to examine
the role of spatial and spin symmetries in the entanglement of a state. 
In this context, we will explicitly discuss the issue of degeneracies, 
noting that the entanglement obtained for different
degenerate states are not the same. Additionally, for both the chain and 
icosahedral systems, a comparative
study will be made between the entanglement profiles (across the 
energy spectrum) and the corresponding profile of the density of 
states (DoS) leading to some interesting observations.

In the next section (Sec. 2), we will first discuss the theoretical 
background of our work where we discuss the measure of bipartite entanglement 
for pure states. Subsequently, we will introduce the different
Hamiltonians used and the numerical techniques employed. In Sec. 3, we 
will present and discuss our results, first for chains and then for the 
icosahedral system. At the end of this section, we will
compare the entanglement profiles with the corresponding DoS profiles. 
We will then conclude our work with some remarks on our observations.

\section{Theoretical Background and Numerical Techniques}

In this section we will briefly describe the entanglement measure, the 
Hamiltonians for strongly correlated systems and the numerical techniques 
used in this work.

\subsection{The Entanglement Measure}

In the introductory section we introduced the notion of entanglement 
intuitively. In a more formal language, the
bipartite entanglement for pure states is defined as follows. Two parts of 
a system are said to be entangled in a particular state if the 
corresponding wave function of the total system cannot be expressed
as a direct product of some states of the two parts. Similar definitions 
can be given for the bipartite entanglement of mixed states and also for the 
multipartite entanglement of both pure
and mixed states, but we will not be interested in these in the present work. 
Quantifying entanglement is generally non-trivial and can have different 
measures according to the usefulness in different contexts 
\cite{amico,horodecki}. Fortunately, bipartite entanglement for pure states 
has a well defined and simple measure, namely, the von Neumann entropy which 
is given by
\begin{eqnarray} S = -Tr ~(\rho~{\rm log_2}~\rho), \label{entrp1} 
\end{eqnarray}
where $\rho$ is the reduced density matrix of either one of the two 
parts, `Tr' implies tracing over an operator (or matrix), and `log' is the 
logarithm which is conventionally calculated with base 2. If $w_i$'s are
the eigenvalues of $\rho$, then $S$ can be written as
\begin{eqnarray} S = -\sum_i ~w_i~{\rm log_2}~w_i. \label{entrp2} 
\end{eqnarray}
Here we note that, $w_i\geq 0$ and $\displaystyle \sum_i w_i = 1$. Now let 
$\displaystyle|\psi\rangle = \sum_{ij} C_{ij} |\phi_i\rangle^l |\phi_j
\rangle^r$ be some pure state (say, an eigenstate of a given Hamiltonian), 
where $|\phi_i\rangle^l$s and
$|\phi_i\rangle^r$s are the normalized basis states of the two parts 
(conveniently called the left and right parts) of a system. Then the elements 
of the reduced density matrix of the left part are given in terms of 
$C_{ij}$ by
\begin{eqnarray} \rho_{ij} = \sum_k C_{ik}C_{jk}^*, \label{rdm}
\end{eqnarray}
with `*' implies complex conjugation. Alternatively, one can consider the
reduced density matrix of the right part whose matrix elements are given by
\begin{eqnarray} \rho'_{ij} = \sum_k C_{ki}C_{kj}^*. \label{rdm1}
\end{eqnarray}
The dimensionalities and hence the number of eigenvalues of $\rho$ and $\rho'$
can be quite different from each other. However,
it can be shown, using the Schmidt singular value decomposition, 
that the {\it non-zero} eigenvalues of $\rho$ and $\rho'$ 
are always equal to each other; hence the von Neumann entropy given in
Eq. (\ref{entrp2}) is the same regardless of whether it is calculated
using the eigenvalues of $\rho$ or $\rho'$.

\subsection{Interacting Model Hamiltonians}

A number of model Hamiltonians for strongly correlated systems will be used 
in this study. The Pariser-Parr-Pople (PPP) model Hamiltonian, which 
includes interactions between distant neighbors in addition
to on-site interactions, is given by
\begin{eqnarray} {\hat H}~=~ -\sum_{<ij>,\sigma}t_{ij}
(\hat c^{\dagger}_{i\sigma} \hat c_{j\sigma} ~+~ H.c.)~+~ \sum_i
\frac{U_i}{2}\hat n_i(\hat n_i-1) ~+~\sum_{i>j}V_{ij}(\hat n_i-z_i) 
(\hat n_j-z_j). \label{ppp} \end{eqnarray}
Here, the first term of the Hamiltonian is the nearest-neighbor transfer 
term also called the H\"uckel term, with $\hat c_{i\sigma}^{~\dagger}$ 
($\hat c_{i\sigma}$) creating (annihilating) an electron with spin 
$\sigma$ at the $i^{th}$ site, and the summation is over a bonded pair of
sites $<ij>$ (in our case nearest neighbors). The second term is the Hubbard 
term with $U_i$ being the on-site repulsion energy for the $i^{th}$ site; here 
$\hat n_i = \sum_\sigma c^{\dagger}_{i\sigma} c_{i\sigma}$ denotes 
the number operator for the $i^{th}$ site. The Hamiltonian with only the 
first term described the one-band tight-binding model or the
H\"uckel model while inclusion of the second term gives the Hubbard 
Hamiltonian \cite{hubbard}. The third term brings in the effects of 
long-range interactions, where $V_{ij}$ is the density-density 
electron repulsion integral between sites $i$ and $j$, $z_i$ is the local
chemical potential given by the occupancy of the $i^{th}$ site for which 
the site is neutral. This term was first introduced for conjugated $\pi$ 
electronic systems, by Pariser and Parr \cite{pariser} as well as by 
Pople \cite{pople} independently in 1953. We employ the Ohno interpolation 
scheme (applicable to conjugated systems) to parametrize $V_{ij}$ \cite{ohno},
\begin{eqnarray}
V_{ij}=14.397\left[\left(\frac{28.794}{U_i+U_j}\right)^2+r_{ij}^2\right]
^{-1/2}. \label{ohno_eq} \end{eqnarray}
Here $r_{ij}$ is the distance (in \AA\hspace*{.09cm}) between the $i^{th}$ and 
$j^{th}$ sites and the energies are in eV. One uses $t = -2.4$ eV and 
$U = 11.26$ eV as the standard parameters for the PPP Hamiltonian of a 
conjugated system.  In the large $|U/t|$ limit,
Hamiltonian in Eq. (\ref{ppp}) reduces to an isotropic spin Hamiltonian 
(the Heisenberg Hamiltonian). The Heisenberg Hamiltonian with site 
spin $\vec{S}_i$ at the $i^{th}$ site can be written as,
\begin{eqnarray} H=\sum_{<ij>}J_{ij} \vec{S}_i\cdot \vec{S}_j,
\label{hspin} \end{eqnarray}
where $J_{ij}$ is the exchange coupling constant between sites $i$ and $j$.

\subsection{Numerical Techniques}

For one-dimensional electronic systems we will use the Rumer-Pauling valence 
bond (VB) basis, which is spin-adapted and complete but non-orthogonal 
\cite{pauling,soos}. The $C_2$ symmetry
and electron-hole symmetry are applied to break the spin adapted space 
(only singlets and triplets are considered in our study) into four symmetry 
adapted subspaces, namely, $A_g^+$,
$A_g^-$, $B_u^+$ and $B_u^-$ \cite{ramasesha,bondeson}. Once we 
diagonalize a Hamiltonian in some
particular space, we get eigenstates which are linear combinations of 
symmetrized VB basis states. These eigenstates can be expressed as linear 
combination of the initial VB basis (unsymmetrized) states. We can also 
expand the VB basis states into linear combination of basis states with a 
constant values of $M_S$ (i.e., $S_z$) \cite{ramasesha,sahoo1}, and can thus 
express the eigenstates as linear combinations of states with a particular 
value of $M_S$. This constant $M_S$ basis conserves the $z$-component of the 
total spin and being orthonormal, it is convenient to use in our study. For 
spin chains (both spin-1/2 and spin-1), we only studied the ground state to 
compare them with those from the electronic models. We also
recently developed a hybrid VB-constant $M_S$ basis technique, which can 
exploit both spin and spatial symmetries of an arbitrary point group 
\cite{sahoo1,sahoo2}. Here the eigenstates are automatically obtained as 
linear combinations of constant $M_S$ basis. We used this hybrid method
for the entanglement study of a half-filled 12-site icosahedral system 
(see Fig. \ref{ico_fig}) which has a huge Hilbert space (with dimension 
1,778,966) \cite{sahoo1}. Once we get an eigenstate as a linear combination 
of constant $M_S$ basis, we can use Eq. (\ref{rdm}) to compute the
reduced density matrices. These matrices can be diagonalized to obtain the 
eigenvalues and subsequently the entropy using Eq. (\ref{entrp2}). In Table 
(\ref{tab_ico}) we show the dimensionalities of different spin and
spatial symmetry adapted subspaces of the icosahedral system. These 
dimensionalities have some significance related to the entropy profiles of 
different subspaces; this will be discussed in Sec. 3.2.

Since all our results will be interpreted in terms of the VB basis, we give 
a very brief description of the basis \cite{soos}. For spin systems, the
spin of a magnetic site ``$s_i$" is replaced by $2s_i$ spin-half objects. 
Now to obtain a state with total spin $S$ from $N$ such
spin-1/2 objects from all the magnetic centers of the system, a total of 
$2S$ such objects are left unpaired and the remaining  $N - 2S$ of the 
spin-1/2 objects are singlet spin paired explicitly, subject to the following 
restrictions: (i) there should be no singlet pairing of any two spin-half
objects belonging to the same magnetic center (this ensures that the $2s_i$ 
objects are in a
totally symmetric combination), (ii) when all the spin-half objects are 
arranged as $N$ dots on a regular $N$-sided polygon and lines are drawn 
between spin paired objects, there should be no intersecting lines in the 
diagram, and (iii) these lines should not enclose any unpaired spin-1/2
objects which are connected by an arrow when all the sites are aligned on a 
straight line and the VB diagram is drawn such that all singlet lines are 
above this straight line. These rules follow from the generalization of the 
Rumer-Pauling rules to objects with spin greater than 1/2 and total spin 
greater than zero. The set of diagrams which obey these rules are called 
``legal" VB diagrams (or basis states). All legal VB diagrams form a complete 
and linearly independent set. VB diagrams which do not follow these rules are 
called ``illegal" and can be expressed as linear combination of ``legal" 
diagrams. Note that for an odd number of spin-1/2 objects and total spin
$S = 1/2$, we cannot have an arrow. In this case we add a $``phantom”$ site 
and generate singlet diagrams; the phantom site does not appear in the 
Hamiltonian.

For electronic systems, a given orbital can be in one of four states; it can 
be (1) empty, (2) singly occupied with an up-spin electron, (3) singly 
occupied with a down-spin electron, and (4) doubly occupied. Let $N$ be the 
number of orbitals, $N_e$ be the number of electrons with $N_{\uparrow}$ 
up-spin electrons and $N_{\downarrow}$ down-spin electrons, so that $N_e = 
N_{\uparrow} + N_{\downarrow}$. Now, for a fixed occupancy of the orbitals, 
a linearly independent and complete set of states with total spin $S$ and 
$M_S = S$ is obtained by the extended Rumer-Pauling rules as follows. 
(i) The $N$ orbitals are arranged as dots along a straight line. 
(ii) Doubly occupied sites are marked as crosses. (iii) An arrow is passed
through $2S$ of the singly occupied sites, passing on or above the straight 
line on which the system is represented. The arrow denotes the spin coupling 
corresponding to total spin $S$ and
total $z$-component $M_S = S$. VB diagrams corresponding to other $M_S$ 
states can be obtained by operating on the arrow with the $S^-$ operator the 
required number of times. (iv) Remaining singly occupied
sites are singlet paired and are denoted by lines drawn between them. 
(v) Diagrams with (a) two or more crossing lines, or (b) crossing line and 
the arrow, or (c) a line enclosing the arrow are called ``illegal" and are 
rejected. The remaining set of diagrams
correspond to a complete and linearly independent set of VB states for the 
chosen orbital occupancy. The set of VB diagrams which obey the extended 
Rumer-Pauling rules are called ``legal" VB diagrams. Note that for $N_e$ odd 
and total spin $S = 1/2$, we cannot have an arrow. As with the
pure spin case, we handle this situation by the concept of a $``phantom"$ site.

\begin{figure}[]
\begin{center} {\includegraphics[width=9.0cm]{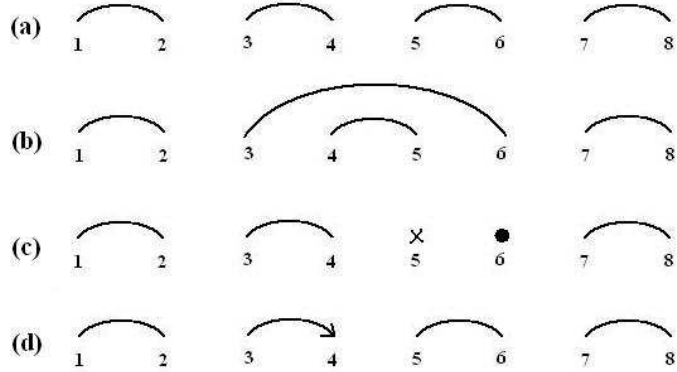}}
\caption{\small Some legal VB diagrams are shown for an 8-site system. The 
first three diagrams, (a), (b) and (c), are for total spin zero while the
last one, (d), is for total spin one. Diagram (c) is possible only for 
electronic systems.} \label{vbbasis} \end{center} \end{figure}

Some legal diagrams are shown in Fig. \ref{vbbasis}; these will be referred 
to in the next section.

Before finishing this section let us stress two important points: (i) the 
procedure above will give us states with $M_S$ equal to the total spin $S$, 
and (ii) any basis (a diagram) can be
written as a direct product of states of lines (representing singlets), 
arrow (representing the spin of the diagram) and crosses (representing doubly 
occupied sites) in any order. Since each of the
objects contain a pair of fermionic operators, those operators commute.
Note, crosses or dots do not appear for pure spin systems. If we calculate 
the entropy for a VB diagrammatic basis, it would give a non-zero value 
only if the boundary plane dividing the system
into two parts (left and right) intersects at least one line of the diagram. 
If the boundary plane intersects only an arrow (if present), it would give 
zero entropy as the arrow is the direct product
state of all the unpaired spins (or electrons) as we have chosen $M_S = S$. 
To clarify this we note that in diagram (a) of Fig. \ref{vbbasis}, if the
boundary plane intersects the system in between sites 3 and 4, the calculated 
entropy will not be zero; on the other hand, when the plane intersects the 
system in between sites 4 and 5, the calculated entropy will be zero. Since the
eigenstates of a system can be written as linear combination of VB diagrams,
the entropy calculated for an eigenstate would be high if the leading VB 
diagrams (with large coefficients) themselves have non-zero entropy.

\section{Results and Discussion}

In this section, we first present results for spin and electronic system for 
the chains. We follow this with results and discussions for the icosahedral 
cluster. In the last part of the section, we
compare and discuss the DoS profiles with the corresponding entanglement 
profiles for different spaces of both the chain and icosahedral systems.

\subsection{Entanglement Studies of Chains}

We have performed entanglement studies on the chain system mainly to 
explore the effect of long-range interactions. In these studies we will also 
see the role of the chain length and block size on the entanglement 
entropy. For this purpose we study Hubbard chains (with $U/t =$ 0, 
2, 4, 6, 8, 12 and 40) and PPP chains. Electronic systems are always taken to 
be half-filled. We also study the entanglement
of spin-1/2 and spin-1 Heisenberg antiferromagnetic spin chains in order to 
compare its size dependence with the results for electronic systems.

\begin{figure}[]
\begin{center} {\includegraphics[width=16.0cm]{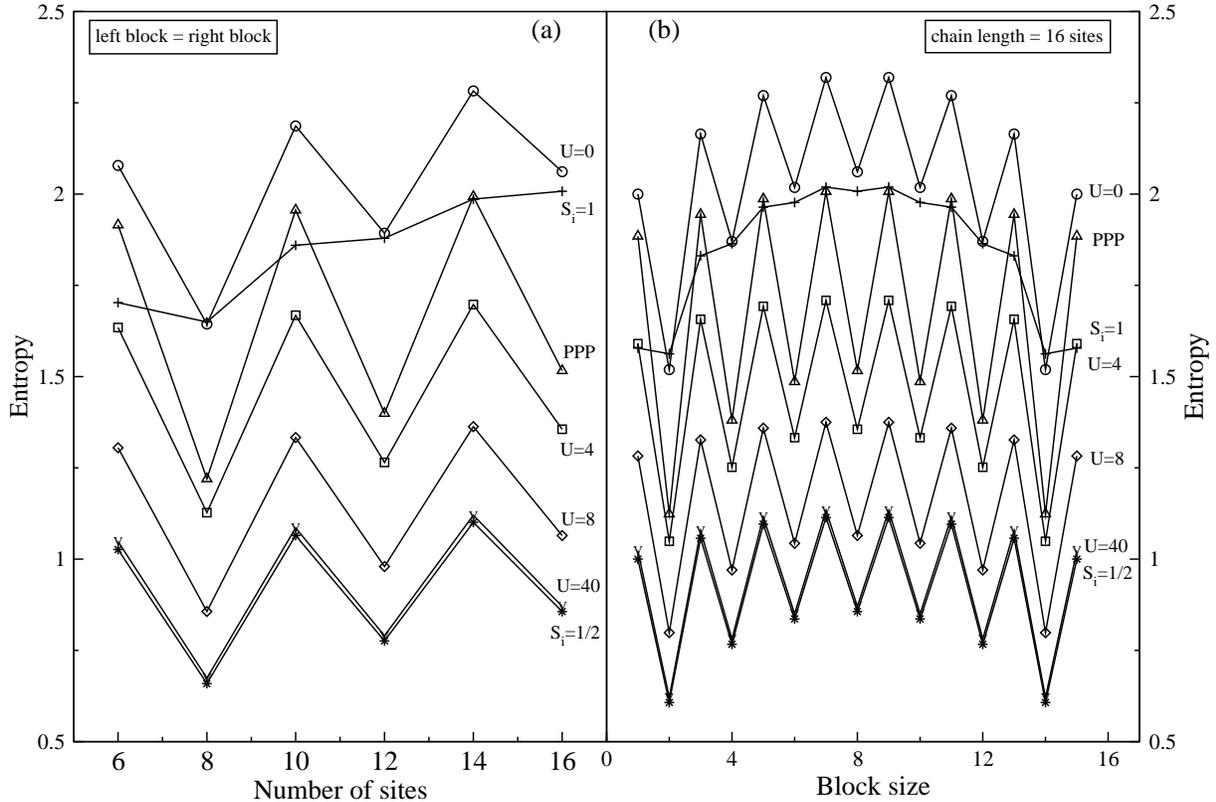}}
\caption{\small (a) Ground state entanglement studied with equal block 
sizes and increasing chain length. (b) Ground state entanglement studied 
with different block sizes for a fixed chain length (16 sites).} 
\label{entr_grndst} \end{center} \end{figure}

First we study the entanglement in the ground state of different models, with 
respect to variations in both the chain length and the block size (see Fig. 
\ref{entr_grndst}). We find that for the electronic
systems, for odd block sizes (having an odd number of sites), the 
entropy is high, while for even block sizes the entropy is low. This can be 
explained from the VB theory. In the ground state of a chain,
the contribution of the Kekule basis states (see Fig. \ref{vbbasis} a), 
is large since this basis state corresponds to nearest-neighbor singlets. 
Now depending on the position of the boundary plane (whether it intersects a 
singlet line or not; see the discussion in Sec. 2.3), the entropy of the ground
state will be high or low. This is the reason why for equal block sizes, the 
$4n+2$ ($n$ being a positive integer) chains have higher entropy than the 
$4n$ chains; in the former, we cut a singlet line in the Kekule diagram while 
in the latter, the system is divided between nearest-neighbor lines of the 
Kekule diagram (see Fig. \ref{entr_grndst} a). In order to further verify 
this odd-even effect, we also studied 16-sites fermionic chains, with 
different sizes of blocks (see Fig. \ref{entr_grndst} b). Here, we find 
that the entropy is large (small) whenever it has an odd
(even) number of sites in a block.

In this regard we also note that the odd-even effect is more prominent or 
sharper for the PPP model compared to that for the Hubbard model. This is 
because the PPP term (the third term in Eq. (\ref{ppp})) is a diagonal 
interaction term between empty and doubly occupied sites. The lines or 
arrow (singly occupied sites) remain unchanged by the term. This signifies 
that the long-range term favours ionic states; as a result, the number of 
covalent states (as in Fig. \ref{vbbasis} b) present in the 
ground state is less compared to the case of the Hubbard model. Now due to 
dominance of diagrams like in Fig. \ref{vbbasis} a among the other less important 
covalent diagrams present in the ground state (anyway ionic diagrams contribute 
less towards entropy due to lack of lines), odd-even effect becomes sharper for 
the PPP model.

We further notice that, within the even or odd block sizes, the entropy 
increases as shown in Fig. \ref{entr_grndst} a and it takes a dome-like 
structure as shown in Fig. \ref{entr_grndst} b. This is because 
the entropy of a system generally increases with the Schmidt number 
(coming from the Schmidt decomposition). As this number grows as the minimum of
the Fock space dimensions of two blocks, in the first case 
(Fig. \ref{entr_grndst} a) it increases linearly with the system size and 
in the second case (Fig. \ref{entr_grndst} b) it is maximum when the block sizes 
are the same.

Another important observation for the Hubbard system is that the entropy 
decreases with increasing $U/t$. This is because an increase in $U/t$ leads 
to a smaller contribution from the ionic functions to the ground state. This 
effectively reduces the number of states involved (or equivalently the
Schmidt number), which results in a decreased entropy. This fact is further 
supported by the result that the entropy profile for the Hubbard chain 
approaches that of the spin-1/2 antiferromagnetic Heisenberg spin chain with 
increase in $U/t$. We remember here that the ground state of the spin
chain is dominated by the Kekule states.

In order to study the entanglement behavior of spin systems and compare the 
result with the corresponding behavior of electronic systems, we have also 
studied the entropy in the ground state of the
spin-1/2 and spin-1 Heisenberg spin chains. Firstly, we note that the 
entropy profile of the spin-1/2 chain closely matches that of the Hubbard 
chain with $U/t$ = 40, and it exhibits a strong
odd-even effect. The entropy of the spin-1 chain, on the other hand, shows 
a very weak odd-even effect. The magnitude of the entropy is higher in all 
cases compared to the spin-1/2 chain due to the large Fock space dimensionality
of the spin-1 chain (note that a spin-1 can be considered as two spin-1/2's 
without the singlet combinations between them). The weakness of the odd-even 
effect in the spin-1 chain implies that not only the Kekule states, but also
other VB basis states (where singlets form
between distant sites) contribute significantly to the ground state.

\begin{figure}[]
\begin{center} {\includegraphics[width=16.0cm]{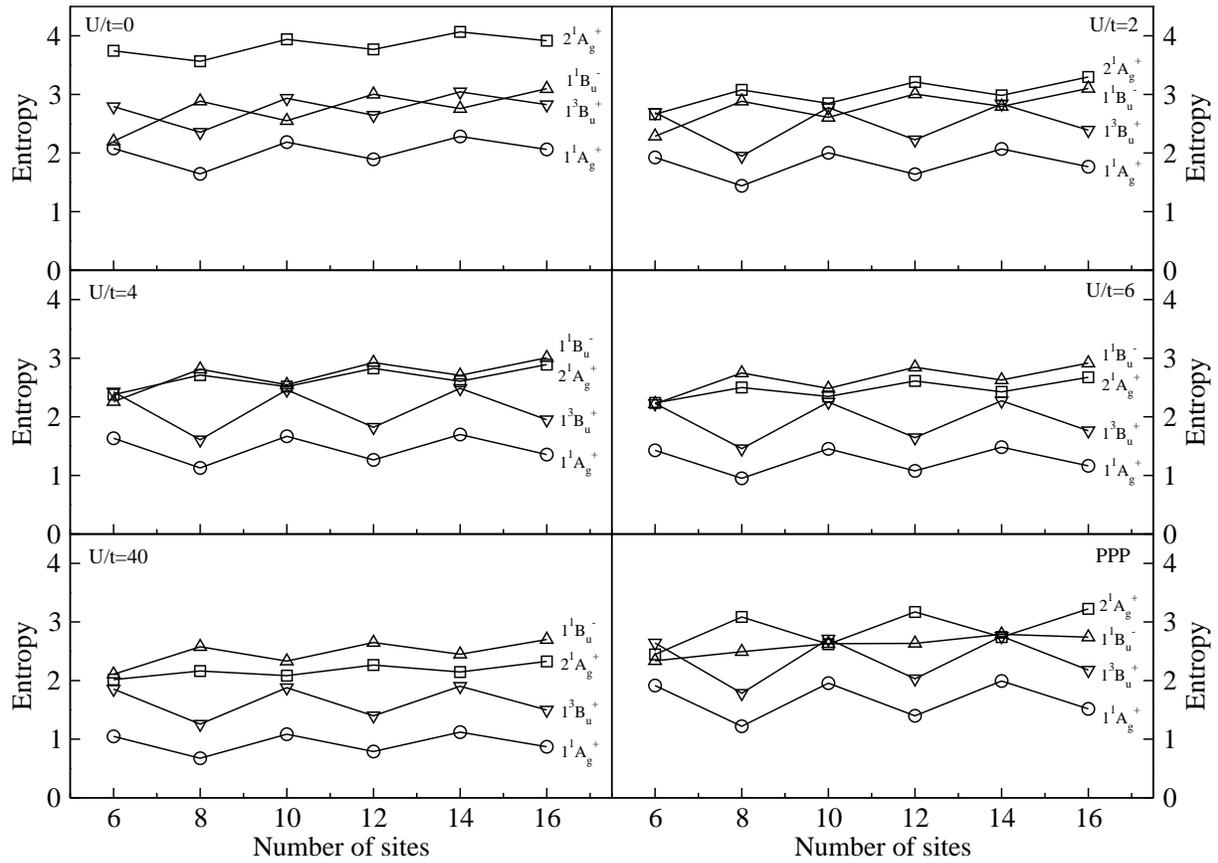}}
\caption{\small Entropy for some important excited states (along with the 
ground state) are shown here as a function of chain length (equal block size). 
Values shown here are for the Hubbard model with $U/t$ = 0, 2, 4, 6, 40, and 
for the PPP model.} \label{entr_excite_hub_ppp} \end{center} \end{figure}

We also study the entanglement for some important excited states of the Hubbard 
model for different $U/t$ values and the PPP model (see Fig. 
\ref{entr_excite_hub_ppp}). Here we find that the entropy for 2$^1A_g^+$ 
(lowest energy two-photon singlet state) is higher compared to the other 
states considered when $U$ =0. This is due to the fact that its energy is 
higher compared to others, hence it has a greater admixture of basis states 
like the ones shown in Fig. \ref{vbbasis} b. Since these basis states 
contain singlet 
lines between distant neighbors, the state has a high entropy. However, the 
entropy of this state decreases as $U$ increases. This is because as $U/t$ 
increases, the 2$^1A_g^+$ state becomes the spin wave excitation of a 
spin-1/2 Heisenberg antiferromagnetic chain. As already
discussed, the entropy of spin chains is smaller than that of the fermionic 
chains. We also find that the odd-even effect in this state is smaller 
compared to the other states in the Hubbard model while the effect
is nearly absent in the $1^1B_u^-$ state of the PPP model. This observation in 
the Hubbard model can be understood from the fact that the $2^1A_g^+$ state is
composed of two triplets. These triplets tend to be separated. This implies 
that in the half-blocks, the favored states correspond to large spin and hence
the odd-even effect will be less pronounced. The absence of the odd-even 
effect in the $1^1B_u^-$ state for the PPP model can also be understood from 
the fact that in the one-photon state, purely covalent
basis states are absent, and that the PPP term favors basis states with more 
ion-pairs. Since there are many such ionic basis states present in
significant amounts in this state, the odd-even effect will be suppressed
here. We further notice that the odd-even effect reverses for $2^1A_g^+$ 
state when one goes from $U = 0$ to even a small value (we have taken 
$U = 2$ here). An understanding of this feature requires further analysis. 

\begin{figure}[]
\begin{center} {\includegraphics[width=16.0cm]{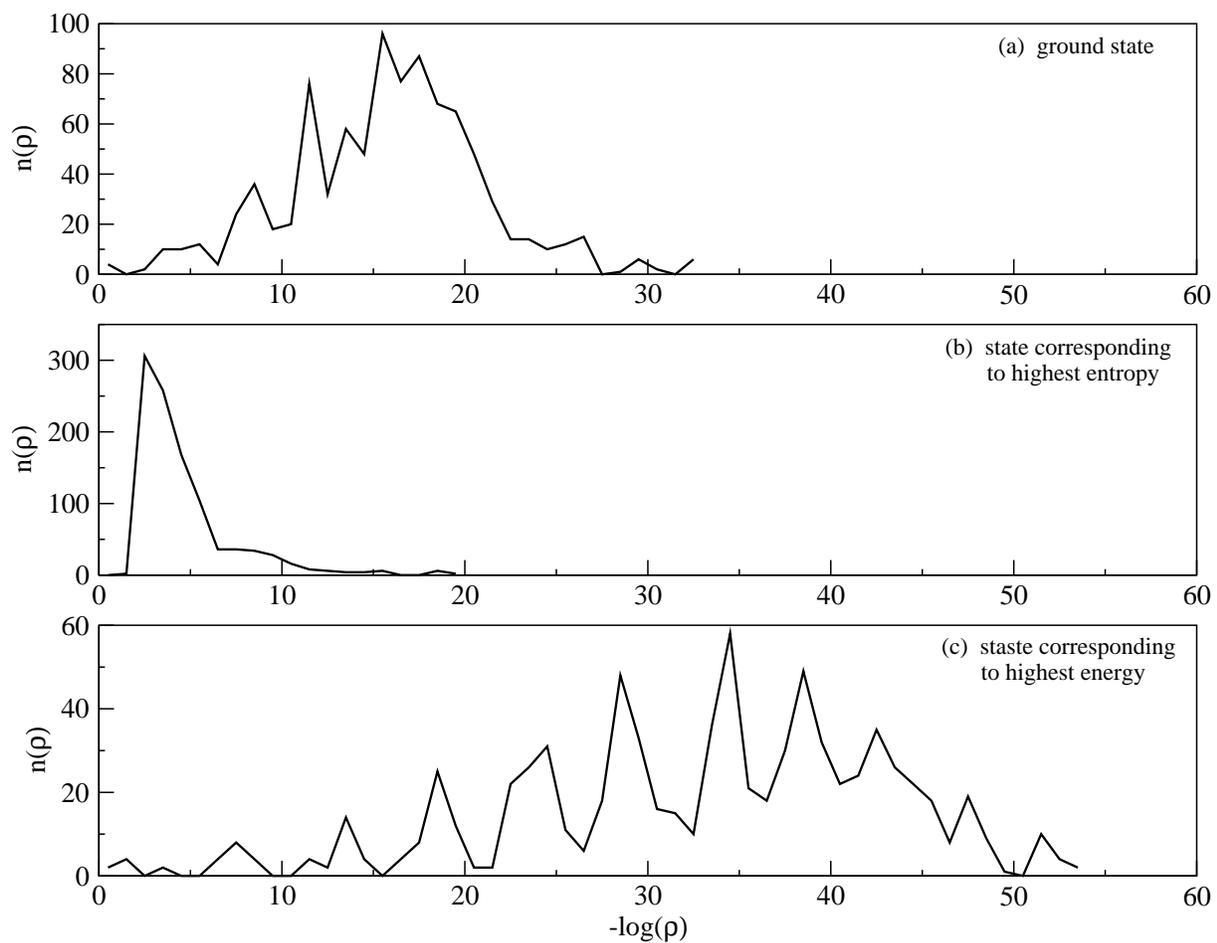}}
\caption{\small DoS profiles of eigenvalues of reduced density matrix (RDM) 
for three different states 
from the $^1A_g^+$ space. The result is obtained for the half-filled 10-site 
chain using the PPP Hamiltonian.} \label{agp_rdmdos} \end{center} \end{figure}

\begin{figure}[]
\begin{center} {\includegraphics[width=16.0cm]{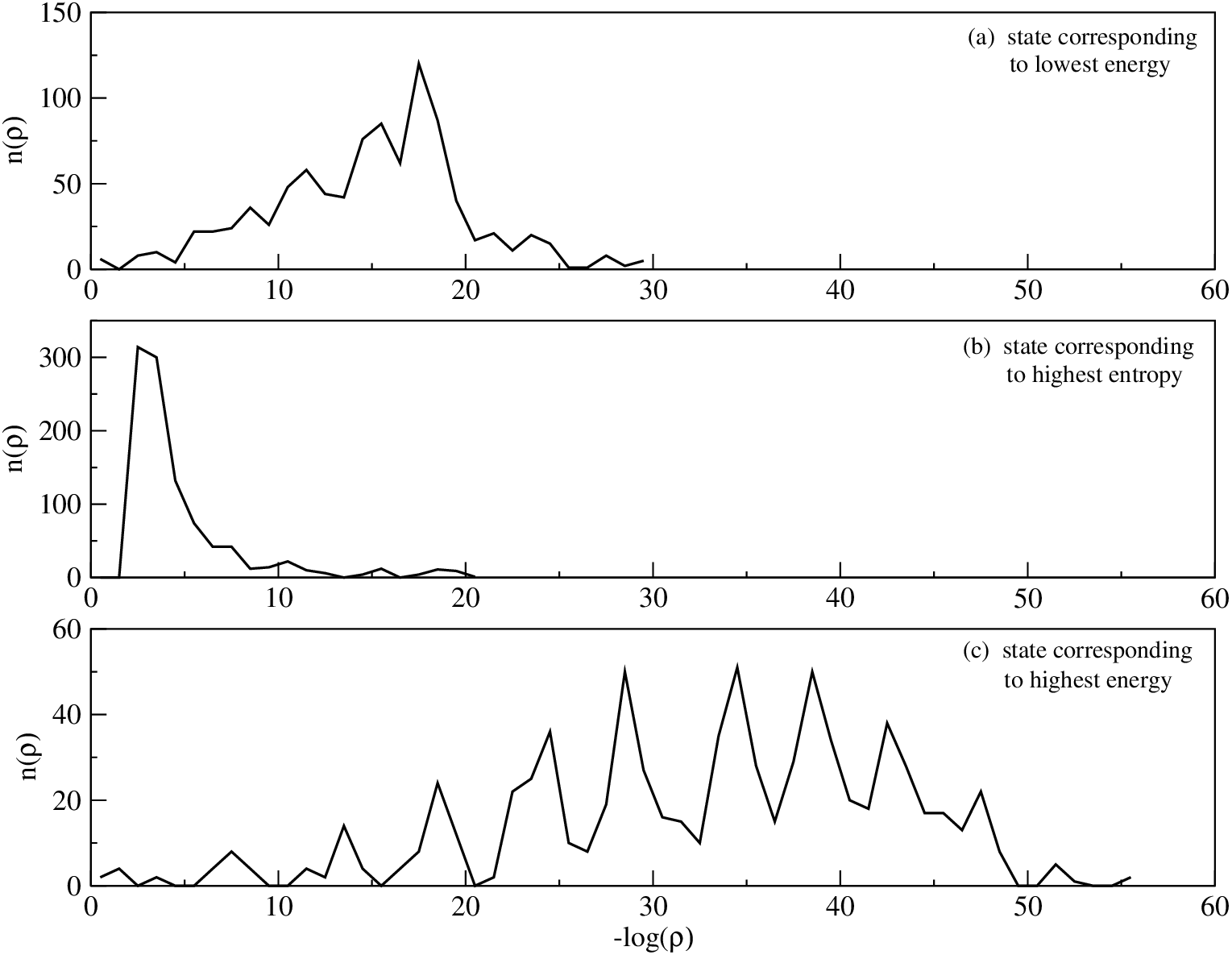}}
\caption{\small DoS profiles of eigenvalues of RDM for three different states 
from the $^1B_u^-$ space.  The result is obtained for the half-filled 10-site 
chain using the PPP Hamiltonian.} \label{bum_rdmdos} \end{center} \end{figure}

\begin{figure}[]
\begin{center} {\includegraphics[width=16.0cm]{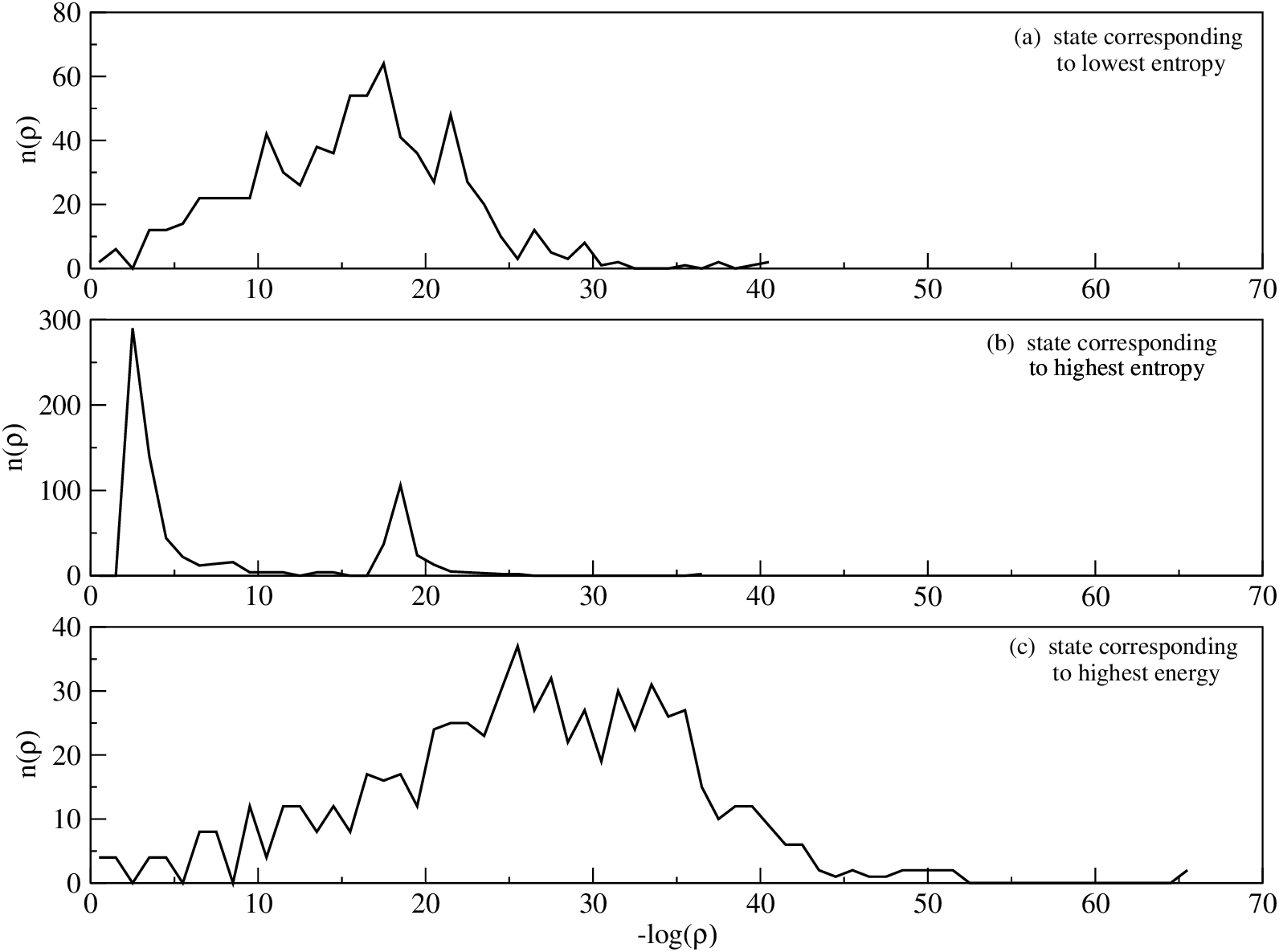}}
\caption{\small DoS profiles of eigenvalues of RDM for three different states 
from the $^3B_u^+$ space. The result is obtained for the half-filled 10-site 
chain using the PPP Hamiltonian.} \label{3bup_rdmdos} \end{center} \end{figure}

To better understand the dome-like structure of the entropy across the energy 
spectrum, we histogram the number $n_p(\rho)$ of density matrix eigenvalues 
in the range $10^{-p} > \rho > 10^{-(p+1)}$ and plot it against the negative 
logarithm of eigenvalues $-{\rm log_2} \rho$. 
These are shown for states from the spaces $1^1A_g^+$, $1^1B_u^-$ and $1^3B_u^+$ 
in Figs. \ref{agp_rdmdos}, \ref{bum_rdmdos} and \ref{3bup_rdmdos} respectively.
{}From the profiles we note that for states with very low and very high 
energies, the histograms have a broader
distribution, while the states in the middle of the energy spectrum
have a narrow distribution (having large $n(\rho)$) and are centered towards 
the origin of the axis. It is clear from Eq. (\ref{entrp2}), that this 
distribution leads to a large entropy for the states with intermediate 
energies and a small entropy for states at either end of 
the energy eigenvalue spectrum.

Our results also show why the DMRG technique is so successful for the 
one-dimensional PPP model even though the model has long-range interactions.
Note that the long-range interactions renders the system topologically 
higher dimensional since cutting the system leads to dangling bonds 
whose number increases with increasing system size; hence, according to the 
area law, the cut-off in the number of DMEVs for a fixed accuracy should 
increase with the system size. But we can see 
from Fig. \ref{entr_grndst} that, even though the odd-even effect is sharper 
for the PPP model compared to different short-range models, the change of 
the entropy with the system size is essentially the same as for
the short-range models. This can be understood in the following way: the 
long-range part of the PPP Hamiltonian in Eq. (\ref{ppp}) involves diagonal 
interactions between sites which are empty or doubly occupied. This term 
only weakly affects the weightage of states with large single occupancies
(i.e., lines and arrow, in the VB language). Since the covalent states (lines)
which actually contribute towards the entropy are not significantly affected 
by this long-range term, the overall entropy of an eigenstate does not change 
much. The only consequence of this term is the sharper odd-even effect which 
results, as we already explained, due to the dominance of diagrams like in 
Fig \ref{vbbasis} a in the ground state of the PPP model. 

\begin{figure}[]
\begin{center} {\includegraphics[width=16.0cm]{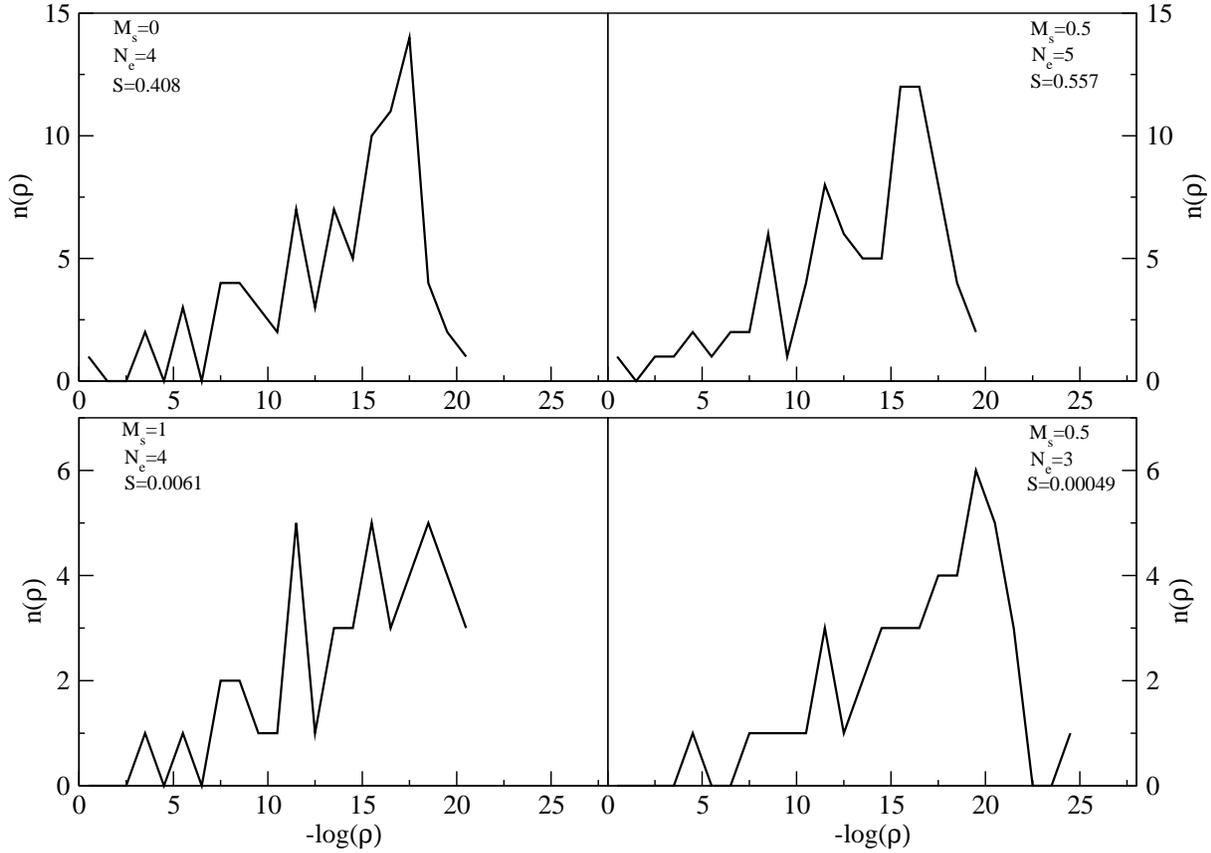}}
\caption{\small DoS profiles of eigenvalues of RDM for different sectors 
corresponding to different electron numbers and $z$-components of total spin. 
The state studied here is the ground state ($A_g^+$) of a 10-site chain using 
the PPP Hamiltonian.} \label{entg_spct_agp} \end{center} \end{figure}

To further substantiate our claim that the long-range interactions do not 
affect the entanglement entropy significantly, we have considered the the 
ground state and a few excited states of a chain of 10 sites in the PPP 
and Hubbard models and calculated the entanglement entropy within different 
sectors of total electron numbers and $z$-component of total spin for one 
half of the system. The idea of an entanglement 
spectrum was introduced in Ref. \cite{li}, and it has proved to be very 
useful for studying a number of strongly correlated quantum systems; in that 
work, the entanglement spectrum was obtained by calculating $-log_2(w_i)$ for 
all the eigenvalues of the reduced 
density matrix in different sectors corresponding to the values of various
additive quantum numbers (i.e., eigenvalues of conserved quantities whose 
values for the entire system are equal to the sum of their values for the 
two parts of the system). In the same spirit, we have separated out the 
contributions of the different total $S_z$ and $n$ sectors, where the former 
corresponds to the $z$ component of the total spin and the latter to the 
number of electrons in the part of the chain for which we compute the 
entanglement entropy; within each of these sectors, we have calculated the 
sum $-\sum_i w_i {\rm log_2} w_i$ (instead of $-{\rm log_2}w_i$ as was done 
in Ref. \cite{li}). The result is shown in Table \ref{en_sp}. In 
Fig. \ref{entg_spct_agp} we show the DoS of the 
reduced density matrix for the PPP model in some sectors of $S_z$ and $n$. 
Due to the electron-hole and spin-inversion symmetries many sectors have the 
same DoS within the numerical accuracy.

Our results show that the entropy for different sectors are almost the same 
for long-range and short-range models. This supports our previous arguments
about the reason for the success of the DMRG method for the PPP model.

\begin{table}[t]
\caption{Entanglement spectrum for different sectors of total electron 
number $N_e$ and $z$-component of total spin $M_s$ for one half of the system. 
We present results for both PPP 
and Hubbard models (with $U/t = 4$) for comparison.}

\label{en_sp}
\centering
\begin{tabular}{|l|c|c|c|c|}
\hline
Model & State & \multicolumn{2}{|c|}{Sector} & Entropy \\
\cline{3-4}
& & M$_{s}$ & N$_{e}$ & \\
\hline
& & 0.5 & 5 & 0.557 \\
& & 0.0 & 4 & 0.408 \\
& A$_{g}$$^{+}$ & 1.0 & 4 & 6.1x$10^{-3}$\\
& & 0.5 & 3 & 4.9x$10^{-4}$\\
& & 1.5 & 5 & 4.6x$10^{-4}$\\
& & 0.0 & 2 & 1.1x$10^{-8}$\\
\cline {2-5}
& & 0.5 & 5 & 0.686 \\
& & 0.0 & 4 & 0.576 \\
PPP & B$_{u}$$^{-}$ & 1.0 & 4 & 1.9x$10^{-2}$\\
& & 0.5 & 3 & 6.1x$10^{-3}$\\
& & 1.5 & 5 & 9.2x$10^{-5}$\\
& & 0.0 & 2 & 6.1x$10^{-7}$\\
\cline {2-5}
& & 0.5 & 5 & 1.049 \\
& & -0.5 & 5 & 0.282 \\
& $^{3}$B$_{u}$$^{+}$ & 1.5 & 5 & 0.282 \\
& & 0.0 & 4 & 0.274 \\
& & 1.0 & 6 & 0.274 \\
& & 0.0 & 6 & 0.274 \\
\hline
& & 0.5 & 5 & 0.545 \\
& & 0.0 & 4 & 0.275 \\
& A$_{g}$$^{+}$ & 1.0 & 4 & 6.1x$10^{-3}$ \\
& & 1.5 & 5 & 1.5x$10^{-3}$\\
& & 0.5 & 3 & 3.5x$10^{-4}$\\
& & 0.0 & 2 & 2.4x$10^{-8}$\\
\cline {2-5}
& & 0.0 & 4 & 0.622 \\
& & 0.5 & 5 & 0.445 \\
Hubbard & B$_{u}$$^{-}$ & 1.0 & 4 & 6.7x$10^{-2}$\\
(U/t~=~4) & & 0.5 & 3 & 3.6x$10^{-2}$ \\
& & 1.5 & 5 & 1.5x$10^{-4}$ \\
& & 0.0 & 2 & 4.2x$10^{-5}$ \\
\cline {2-5}
& & 0.5 & 5 & 1.057 \\
& & 1.5 & 5 & 0.345 \\
& $^{3}$B$_{u}$$^{+}$ & -0.5 & 5 & 0.345 \\
& & 0.0 & 4 & 0.178\\
& & 1.0 & 6 & 0.178\\
& & 0.0 & 6 & 0.178\\
\hline
\end{tabular}
\end{table}

\subsection{Entanglement Studies of an Icosahedron}

To learn if symmetries play any recognizable role in entanglement, we have 
studied a 12-site half-filled icosahedron with the Hamiltonian of the PPP 
model \cite{sahoo1}. The first important issue here is of the degeneracies. 
In general, the entanglement is not the same for degenerate states, and 
a linear combination of two such states can give another degenerate state 
with a different entanglement. In the following we address this problem of 
non-unique entanglement of degenerate states.

\begin{table}[htbp]
\caption{\small Dimensionalities of different symmetry and spin subspaces of
a half-filled icosahedral cluster.}
\label{tab_ico}
\begin{center}
\begin{tabular}{|l||r|r|r|r|r|r|r|} \hline
S$_{tot} \rightarrow$ & & & & & & & \\
$\Gamma$ $\downarrow$ & 0 & 1 & 2 & 3 & 4 & 5 & 6 \\ \hline \hline
A$_{g}$ & 2040 & 3128 & 1684 & 382 & 38 & 3 & 1 \\ \hline
T$_{1g}$ & 16602 & 28821 & 14625 & 3261 & 309 & 6 & 0 \\ \hline
T$_{2g}$ & 16602 & 28821 & 14625 & 3261 & 309 & 6 & 0 \\ \hline
G$_{g}$ & 30272 & 50932 & 26236 & 5880 & 568 & 16 & 0 \\ \hline
H$_{g}$ & 47940 & 79305 & 41255 & 9220 & 900 & 40 & 0 \\ \hline
A$_{u}$ & 1852 & 3188 & 1644 & 348 & 40 & 0 & 0 \\ \hline
T$_{1u}$ & 17082 & 28686 & 14700 & 3372 & 294 & 18 & 0 \\ \hline
T$_{2u}$ & 17082 & 28686 & 14700 & 3372 & 294 & 18 & 0 \\ \hline
G$_{u}$ & 30160 & 50992 & 26176 & 5888 & 560 & 16 & 0 \\ \hline
H$_{u}$ & 46880 & 79680 & 40980 & 9060 & 900 & 20 & 0 \\ \hline \hline
Tot Dim $\rightarrow$ & 226512 & 382239 & 196625 & 44044 & 4212 & 143 & 1 \\
\hline
\end{tabular}
\end{center}
\end{table}

\begin{figure}[]
\begin{center} {\includegraphics[width=5.0cm]{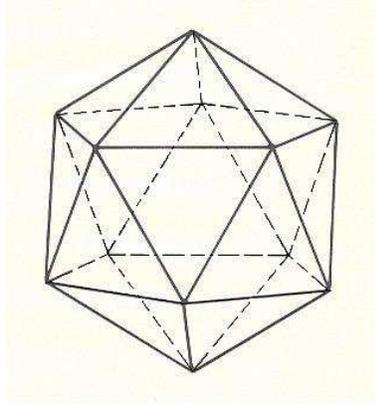}}
\caption{\small An icosahedron system of 12 sites is shown here. The upper six 
sites of the system constitute one part and the rest of the sites constitute 
the other part for our entanglement calculations.} \label{ico_fig} \end{center}
\end{figure}

For spin degeneracies (i.e., same total spin but different $M_S$ values) the 
problem is not severe. Since the $z$-component of the total spin is an 
observable, we can always choose one of its eigenstates. In our study we 
have chosen states with $M_S=S$.

The non-uniqueness problem cannot be trivially solved for the case of 
discrete symmetries. In this case we do not have any observable to select 
one of the degenerate states for our study. It is known that
the projection operator to the full degenerate space is invariant. If 
$\rho_i$ is the density matrix of the $i^{th}$ state in a degenerate manifold, 
then the density matrix averaged over the $g$-fold degeneracy is given by
$\rho_{av} = (1/g) ~\sum_{i=1}^g ~\rho_i$. This $\rho_{av}$ is invariant 
under a change of the basis states. To obtain the reduced density matrix, 
and subsequently the entanglement entropy, we take the upper half of the 
system (upper 6 sites of the Fig. \ref{ico_fig}) as one part and the rest 
as the other part.

\begin{figure}[]
\begin{center} {\includegraphics[width=16.0cm]{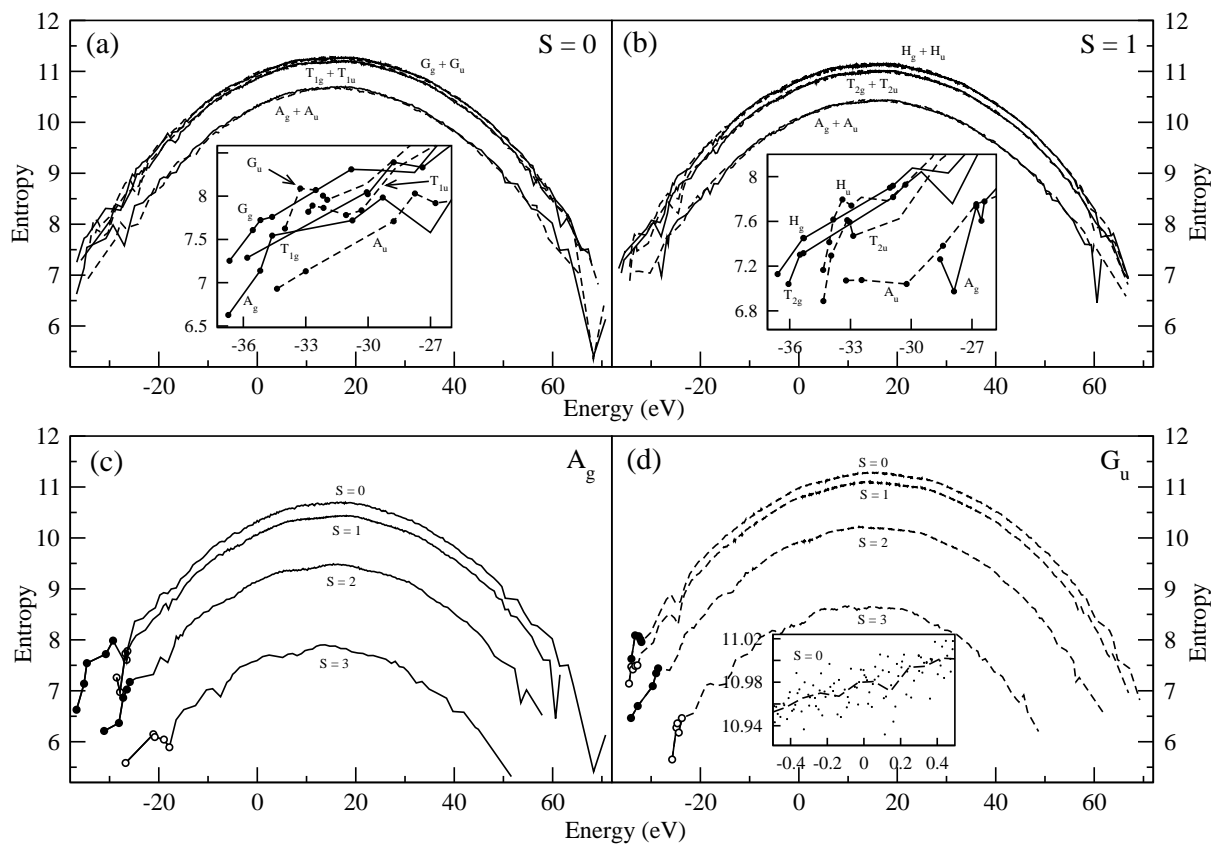}}
\caption{\small Entropy calculated across the energy spectrum of different 
subspaces. Insets of (a) and (b) show entropies (black dots) of the first few 
states of the corresponding subspaces. (c) and (d) show entropies of the 
first few states using filled circles (for even spins) and open 
circles (for odd spins). Inset of (d) shows how entropies fluctuate about a 
mean value. Solid lines represent plots for gerade ($g$) spaces and broken 
lines represent plots for ungerade ($u$) spaces.} 
\label{entr_ico} 
\end{center} 
\end{figure}

The eigenstates of the Hamiltonian correspond to integer spins lying between 
0 and 6 and belong to one of the ten irreducible representations of the 
icosahedral cluster. To see the role of symmetry, we calculate the entropy 
in different spatial symmetry subspaces for two
chosen spin states 0 and 1. This is shown in Figs. \ref{entr_ico} a and b. 
To see the effect of the spin, we calculate the entropy 
in different spin subspaces for two chosen spatial symmetries $A_g$ and $G_u$.
This is shown in Figs. \ref{entr_ico} c and d.

Since the entropy profiles fluctuate (see inset of Fig. \ref{entr_ico} d), 
we take averages to analyze the general features. The averaging is done in 
the following way: the first and last few states in the energy spectrum are 
comparatively well-separated from each other, we leave five of them 
from both sides with their actual values (which are zoomed in the insets of 
Fig. \ref{entr_ico} a and b). Since the rest of the points from both 
sides of the spectrum are closely placed, we average the entropy over sets 
of four consecutive states. This is done for up to 40 states from both 
ends of the spectrum. Since the middle part of the spectrum is highly dense 
\cite{note1}, we average the entropy values over 10 consecutive states.
We find that the entropies for the lowest states in different symmetrized 
subspaces, for a given spin, do not seem to have any special features 
(see insets of Figs. \ref{entr_ico} a and b). However, for a fixed spin, 
the maxima of the entropy profiles (see the middle portions of Figs. 
\ref{entr_ico} a and b) are higher for degenerate spaces, such as $T$, 
$G$, and $H$ compared to the non-degenerate $A$ spaces. In fact, the higher 
the degeneracy in a space, the higher is the entropy maximum; 
for example, the $H$ space has a higher maximum entropy compared to, say, 
the $T$ space. This fact can be partially explained by the fact
that spaces with a higher maximum entropy have a higher dimensionality (see 
Table \ref{tab_ico}). But this does not explain why the entropy of a 
non-degenerate space is much lower compared to the degenerate ones,
while the degenerate ones are very close in entropy. Its explanation lies in 
the fact that entropy increases with DoS (see Sec 3.3) and DoS is much lower 
for non-degenerate spaces \cite{sahoo1}. We also note that for 
a subspace with a particular spatial symmetry, the maxima of the entropy 
profiles is higher for lower spin spaces (see Figs. \ref{entr_ico} c and d). 
This is also generally true for the entropy of the lowest states of different 
spins in a given spatial symmetry subspace. This can be qualitatively explained 
by the dimensionalities of the spaces. However, there must be other factors
contributing to this observation since, in general, the entropy for triplet 
states is lower than that of the singlet states even though the dimension of 
the triplet space is higher than that
of the singlet space. This fact can be explained by noting that the 
low-spin states are made up of more singlet pairs (lines) than the 
high-spin states, and we are only considering states with the highest
$M_S$ value (i.e., $S$) which are least entangled. This leads to a higher 
entropy for states with a lower total spin. If we had worked with the
$M_S =0$ state of the triplet states (which is a linear combination of
two configurations), we would have expected a higher entropy for the states.

The entropy profiles are not monotonically increasing in each symmetry 
subspace; each profile reaches a maxima and then decreases. This can be 
explained by considering the properties of the VB basis. Initially, for the
low-energy eigenstates, the VB states like Fig. \ref{vbbasis} a appear with 
larger weights. This is because these VB states have nearest-neighbor 
singlets. These VB states would contribute less towards the entropy of an 
eigenstate, as the boundary plane crosses a smaller number of lines on the
average. In higher energy eigenstates, more and more VB states, like 
Fig. \ref{vbbasis} b where distant sites form singlets, acquire a significant
weight in the eigenstates. Now, as more states get involved, the entropy of 
the eigenstates starts to increase since the type (b) VB basis states have a 
larger entropy (since the boundary plane crosses a larger number of lines on 
the average). As we move to even higher energy eigenstates, the entropy 
saturates (reaches a maxima) due to the finite number of basis states, since 
we are working 
with a finite system. If we move to the highest energy part of a spectrum, 
more and more ionic VB states like Fig. \ref{vbbasis} c appear and
the type (b) VB states reduce in number. Now due to the reduced number of 
the covalent basis states involved and as ionic VB basis states contribute 
less towards the entropy, the entropy of these states starts to decrease.

\subsection{Comparison of Entropy and DoS profiles}

\begin{figure}[]
\begin{center} {\includegraphics[width=16.0cm]{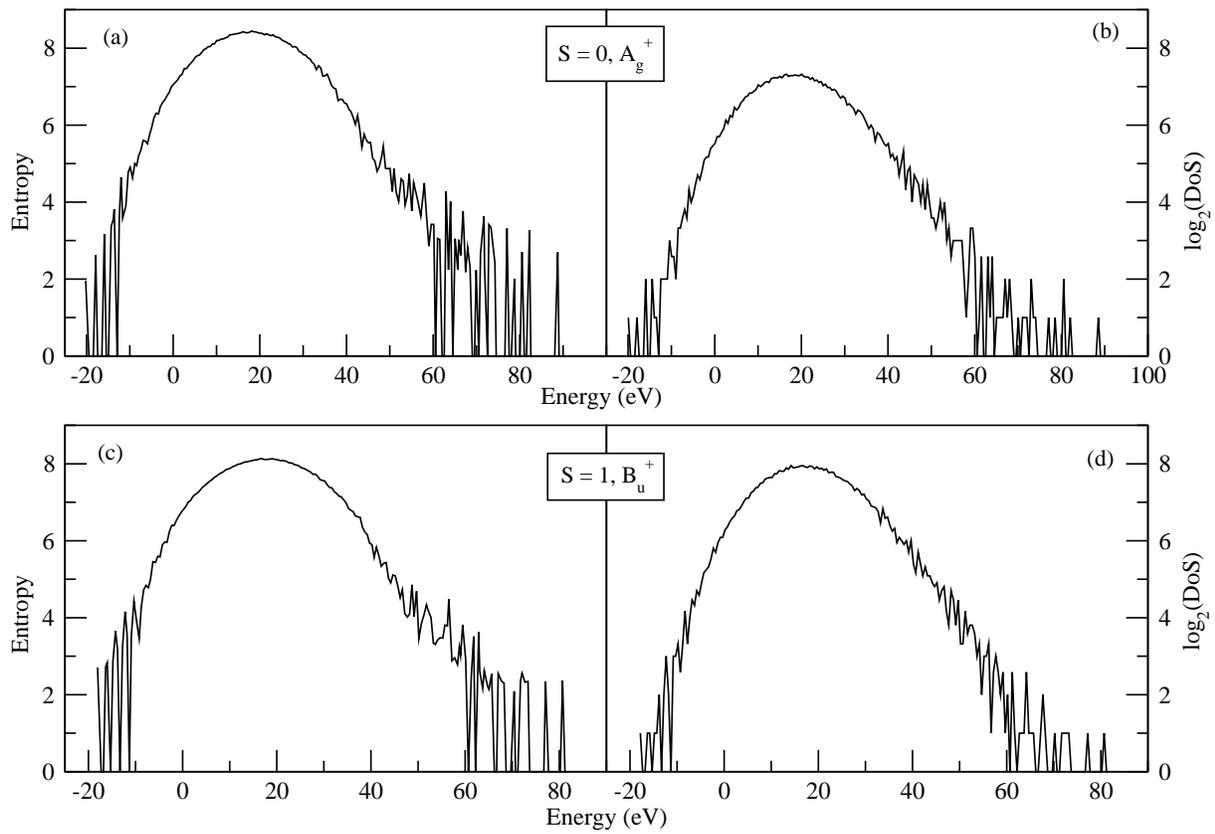}}
\caption{\small For two spin and spatial symmetry adapted subspaces we compare 
the entropy profiles with the logarithm of DoS profiles. The system is a 
10-site half-filled chain.} \label{entr_dos_chain} \end{center} \end{figure}

\begin{figure}[]
\begin{center}
{\includegraphics[width=16.0cm]{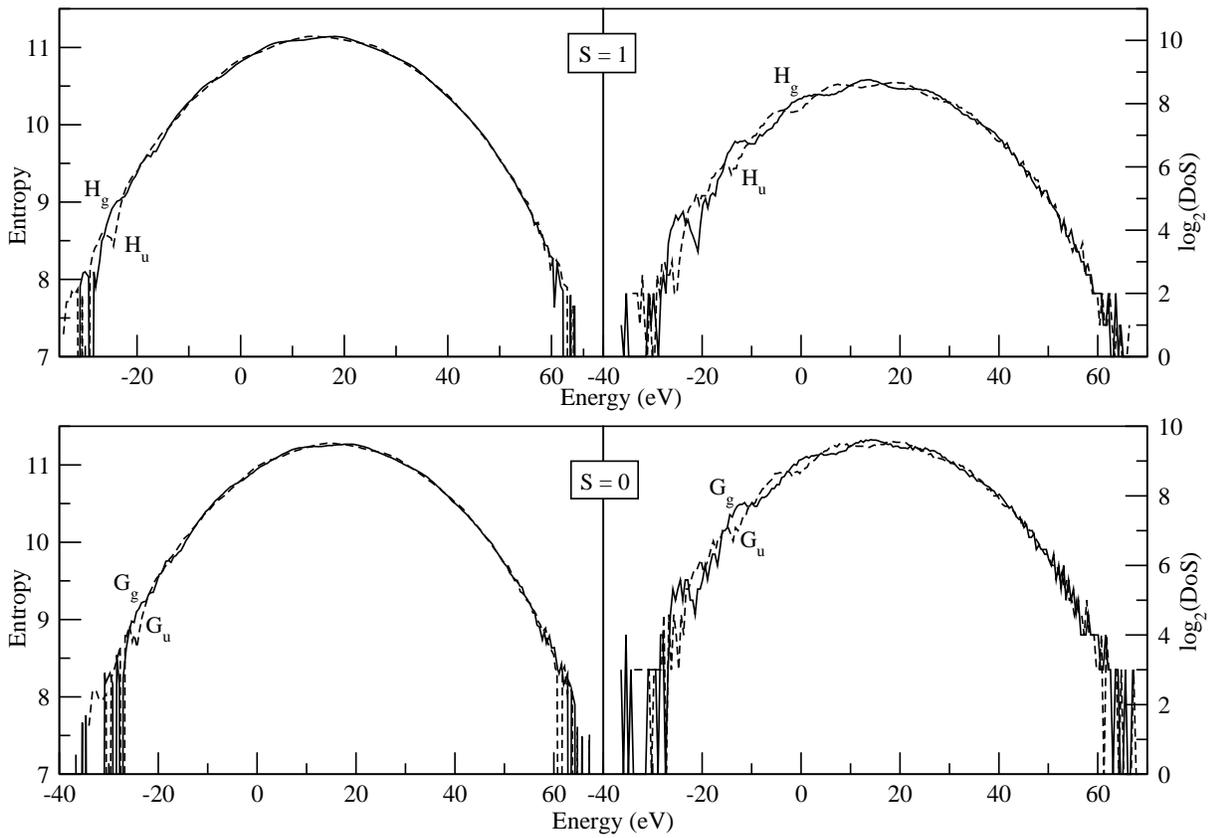}}
\caption{\small For four spin and spatial symmetry adapted subspaces we 
compare the entropy profiles with the logarithm of DoS profiles. The system is
a 12-site half-filled icosahedron.} \label{entr_dos_ico} \end{center} 
\end{figure}

We note that the entropy profiles and corresponding DoS profiles within a 
given symmetry and spin subspace bear an interesting similarity. Using the PPP 
Hamiltonian, we first calculate the eigenvalues and eigenvectors for a given 
spin and spatial symmetry adapted space. We then calculate the entropy 
corresponding to each eigenvector in that space. Using this data, we plot the 
entropy and logarithm of the DoS across the energy spectrum for a few spaces. 
For a chain the results can be seen in Fig. \ref{entr_dos_chain}, while 
for the icosahedron the results can be seen in Fig. \ref{entr_dos_ico}. In 
this study the DoS is calculated by histogramming a number of states within 
an energy gap of $0.5$ eV. To reduce the fluctuation in the entropy profiles, 
we have averaged the entropy of the states in successive energy gaps of $0.5$ 
eV. The entropy is seen to be proportional to the logarithm of the DoS, i.e., 
$S \propto log (DoS)$. Qualitatively, the reason for this can be that 
the higher the DoS, the larger is the number of basis 
states which have a significant weight for a given state in this
energy range, hence, the higher is the entropy. Since the entropy scales as 
the logarithm of the number of states involved, the observed relation
between the entropy and DoS profiles follows.

\section{Conclusion}
To summarize, we have studied the effect of long-range interactions on 
the entanglement of strongly correlated systems. We find that the diagonal 
interactions do not increase the entropy of the states, even if they are 
long-ranged. The behavior of the entropy in correlated 
models can be understood from a VB picture of the eigenstates. The even-odd 
alternation in the entanglement entropy for even/odd block sizes can also 
be understood based on such a VB picture and was illustrated by
analyzing the spin-1 Heisenberg antiferromagnet. We also studied the
effects of spin and spacial symmetry on the entropy of states by examining 
the correlated icosahedral cluster. We showed an interesting similarity 
between the profiles of the entropy and the corresponding profiles of 
the logarithm of the DoS.

\section{Acknowledgment}
S. R. is thankful to the Department of Science and Technology (DST), India for 
financial support and Professor Steve White for a helpful discussion. D. S. 
thanks DST, India for support under SR/S2/JCB-44/2010.



\begin{thebibliography}{}
\bibitem{note1} This can be understood from Fig. \ref{entr_dos_ico}, where 
the DoS is high in the middle parts; also note that there are thousands of 
states in each subspace; for instance, the dimension of $H_g$ with $S=1$ is 
15861 with each state having 5-fold spatial and 3-fold spin degeneracies.
\end{thebibliography}
\end{document}